 \documentclass[12pt]{iopart}
 \usepackage{iopams}
\usepackage{graphicx}
\usepackage{color}

\def\beq{\begin{equation}}
\def\ee{\end{equation}}
\def\bi{\begin {itemize}}
\def\ei{\end{itemize}}

\def\lsim
{\protect \raisebox{-0.75ex}[-1.5ex]{$\;\stackrel{<}{\sim}\;$}}
\def\gsim
{\protect \raisebox{-0.75ex}[-1.5ex]{$\;\stackrel{>}{\sim}\;$}}
\def\lsimeq
{\protect \raisebox{-0.75ex}[-1.5ex]{$\;\stackrel{<}{\simeq}\;$}}
\def\gsimeq
{\protect \raisebox{-0.75ex}[-1.5ex]{$\;\stackrel{>}{\simeq}\;$}}

\def\beq{\begin{equation}}
\def\ee{\end{equation}}

\def\bi{\begin {itemize}}
\def\ei{\end{itemize}}

\def\N{NESS }

\begin{document}

\title{Efficiency of a Brownian information machine}
\author{
Michael Bauer, David Abreu and Udo Seifert
}

\address{
{II.} Institut f\"ur Theoretische Physik, Universit\"at Stuttgart,
  70550 Stuttgart, Germany}
\begin{abstract}
A Brownian information machine extracts work from a heat bath through
a feedback process that exploits the information acquired in a
measurement. For the paradigmatic case of a particle trapped in a
harmonic potential, we determine how power and efficiency for two variants
of such a machine operating cyclically depend on the cycle time and
the precision of the positional measurements. Controlling only  
the center of 
the trap leads to a machine that has zero efficiency at maximum power
whereas additional optimal control of the stiffness of the trap leads to an
efficiency bounded between 1/2, which holds for maximum power, and 1
reached even for finite cycle time in the limit of perfect measurements.

\end{abstract}

\maketitle

\def\lsim
{\protect \raisebox{-0.75ex}[-1.5ex]{$\;\stackrel{<}{\sim}\;$}}

\def\gsim
{\protect \raisebox{-0.75ex}[-1.5ex]{$\;\stackrel{>}{\sim}\;$}}

\def\lsimeq
{\protect \raisebox{-0.75ex}[-1.5ex]{$\;\stackrel{<}{\simeq}\;$}}

\def\gsimeq
{\protect \raisebox{-0.75ex}[-1.5ex]{$\;\stackrel{>}{\simeq}\;$}}

\def\bim{{b_i^-}}
\def\bip{{b_i^+}}
\def\yims{{(y_i^-)}^2}
\def\yips{{(y_i^+)}^2}
\def\N{{\cal N}}
\def\i{{\scriptstyle {\cal I }}} 
\def\I{{\cal I}}

In Kelvin's formulation, the second law of thermodynamics
states that no
work can be extracted from a thermally equilibrated
system through a cyclic process that leaves no trace
elsewhere. If, however, more detailed information of
the system becomes available through a measurement,
then one can indeed extract work as illustrated a 
long time ago with the gedankenexperiments of Maxwell's 
demon and 
Szilard's engine \cite{leff03}. More recently, by combining concepts
from stochastic thermodynamics with those from
information theory, a quantitative
framework has emerged leading to bounds refining
the second law to such feedback driven processes
\cite{touc00,kawa07,cao09,saga10,horo10,espo11,abre11a,lahi12,saga12}.
Specialized to one cyclic process starting in equilibrium, 
the bound
\beq
W\leq \I
\label{eq:bound1}
\ee 
connects the mean extractable work $W$ to
the mean information $\I$ (defined more precisely below)
acquired through the measurement. Brownian particles in 
time-dependent potentials provide a paradigm for such
systems both in recent experiments \cite{toya10a,beru12} as 
in several theoretical case studies \cite{dill09,abre11,horo11a}. 
The latter works have demonstrated
that saturating the bound in (\ref{eq:bound1}) typically requires both an 
infinite cycle time and a sufficient number of control 
parameters in the potential.

The purpose of the present paper is to study these processes
from a perspective that focuses on the performance of such
Brownian information machines in a steady state where
measurements and subsequent optimal driving based on these
are repeated with a finite cycle time $t$. On average per 
cycle, by exploiting the information $\I_*$, the machine extracts the
work $W_*$ thus delivering 
a power $P\equiv W_*/t$. The extant generalization of the bound 
(\ref{eq:bound1}) to such a cyclic operation \cite{abre11a} then motivates 
to define efficiency as
\beq
\eta \equiv W/ \I  
\ee 
following in spirit an earlier approach \cite{cao09}.
Apart from maximum efficiency, it is particularly interesting
to determine efficiency at maximum power.
The later concept has been studied extensively for non-feedback 
driven heat engines operating between two heat baths, see, 
e.g., \cite{schm08,espo09,tu08,blic12} and references therein, and, more 
recently, also for autonomous isothermal machines \cite{seif11a}.

The solution to this problem of performance at a finite cycle time 
cannot trivially be inferred from available results \cite {abre11}
on the 
maximal extractable work following {\it one} measurement in
finite time since at the beginning of the second (and any
further) cycle the system will typically not have reached 
thermal equilibrium again. In fact, the initial state of the
$i-$th-cycle will depend on the result of all previous
measurements which makes the present problem non-trivial.

Our system consists of an overdamped Brownian particle in a harmonic potential
\beq
V(x,\tau)=k(\tau)[x-\lambda(\tau)]^2/2
\ee
with external time-dependent control of the center, $\lambda(\tau)$, and  stiffness,
$k(\tau)$, of the trap \cite {abre11}. Throughout the paper, we use dimensionless variables.
The  harmonic potential has the advantage
that a Gaussian distribution
\beq
p(x)=\N_x(b,y^2)\equiv\frac{1}{(2\pi)^{1/2}y}\exp\left(-\frac{(x-b)^2}{2y^2}\right)\ee
remains Gaussian both under the stochastic dynamics in the
potential and under positional measurements with an error $\pm~y_m$.
The dynamics of the mean $b(\tau)$ and  variance $y^2(\tau)$ of $x$ follows from the
corresponding Fokker-Planck equation as \cite {abre11}
\beq
\dot b(\tau) =k(\tau)[\lambda(\tau)-b(\tau)]
\ee
and
\beq
\dot y(\tau) = y(\tau)[1/y^2(\tau)-k(\tau)] 
\label{eq:y-dyn}
\ee where we denote time-derivatives with a dot throughout.

We now implement a cyclic feedback scheme based on measurements of the position
repeated periodically in intervals of lengths $t$.
At the beginning of the $i-$th cycle, we measure the position $X_i$ with a precision $\pm~ y_m$ 
leading to
the distribution
\beq
p(X_i)=\N_{X_i}(\bim,\yims+ y_m^2)
\label{eq:pX}
\ee for the measured value 
if the distribution prior to the measurement is
characterized by 
\beq
p_i^-(x) =\N_x(\bim,\yims).
\label{eq:pim}
\ee
 After the
measurement the distribution for $x$ follows from Bayes' theorem as
\beq
p(x|X_i)=\N_x(\bip,\yips),
\label{eq:pxX}
\ee
with
\beq
\bip=\frac{X_i\yims+\bim y_m^2}{\yims+y_m^2}
\label{eq:bip}
\ee
and
\beq
\yips= \frac{\yims y_m^2}{\yims+y_m^2}.
\label{eq:yips}
\ee
Based on this measurement, we maximize the
extracted work by optimally adjusting the control parameters.
Quite generally, given an initial state ${b(0)=\bip,y^2(0)=\yips}$ and
a time-dependent $b(\tau)$ and $y(\tau)$, the
extracted work after a time $t$ becomes \cite {abre11}
\beq
W^{\rm out}= W^{\rm out}_b + W^{\rm out}_y
\ee
with
\beq
 -W^{\rm out}_b \equiv [b^2(t)-(\bip)^2]/2 + \int_0^td\tau~\dot b^2(\tau)
\label{eq:Wb}
\ee
and 
\beq
-W^{\rm out}_y \equiv [y^2(t)-\yips]/2 -\ln[y(t)/y_i^+]+ \int_0^t d\tau~\dot y^2(\tau) .
\label{eq:Wy}
\ee
Here, we have required that the trap is centered at $\lambda=0$ with stiffness $k=1$
at beginning and end of the cycle allowing for jumps of these two control parameters.
Depending on the amount of control available, two cases must be distinguished.

\begin{figure}[h]
 \centering
\includegraphics[width=0.45\textwidth]{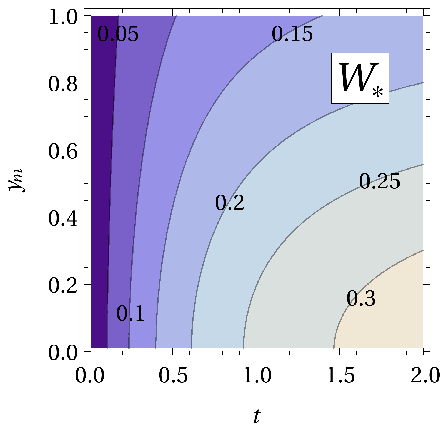}
\includegraphics[width=0.45\textwidth]{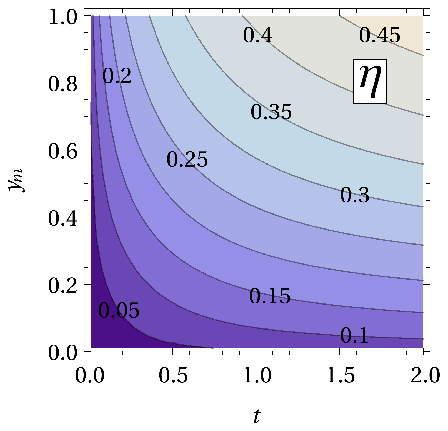}
\caption{Performance of the machine with constant stiffness $k=1$ and optimally controlled center $\lambda(\tau)$.
Extracted work $W_*$ and efficiency $\eta$ both as function of cycle time $t$ and measurement error $y_m$.}
\label{fig:w_k_fest}
\label{fig:eta_k_fest}
\end{figure}

If the stiffness is fixed, $k(\tau)\equiv 1$, only the center of the trap $\lambda(\tau)$ is
controllable. Using (\ref{eq:y-dyn}) in the integral (\ref{eq:Wy}) shows that in this
case $W^{\rm out}_y\equiv 0$. The $b$-dependent term is maximized by a linear
function $b(\tau)=\bip[1-\tau/(2+t)]$ leading to the optimal extracted work
in the $i$-th cycle
\beq
W_i^{\rm out}=(\bip)^2t/[2(2+t)]  .
\ee This work still depends on the result of all measurements $\{X_j\}_{1\leq j\leq i}$.
Conditionally averaging this work over the last measurement will lead to a useful
recursion relation as follows. With
\beq
\langle (\bip)^2 \rangle_{{X_i}} \equiv \int dX_i~ (\bip)^2 p(X_i)
\ee
and (\ref{eq:pX}), (\ref{eq:bip}) and (\ref{eq:yips}) we get
\beq
\langle (\bip)^2 \rangle_{{X_i}} =\frac{4}{(2+t)^2}{(b_{i-1}^+})^2 + \yims-\yips
\ee
and hence
\beq
\langle W_i^{\rm out} \rangle_{{X_i}} =\frac{4}{(2+t)^2}W_{i-1}^{\rm out} + \frac{t}{2(2+t)} \left( \yims-\yips \right) .
\ee
Since the last term is independent of the outcomes of  measurements, subsequent averaging over all
previous measurements $\{X_j\}_{1\leq j\leq i-1}$
(indicated by an unconstrained bracket $\langle ...\rangle$) leads to
\beq
\langle W_i^{\rm out}\rangle  =\frac{4}{(2+t)^2} \langle W_{i-1}^{\rm out} \rangle +  \frac{t}{2(2+t)}\left( \yims-\yips \right) .
\ee
Solving this recursion in the stationary limit, $i\to\infty$, we thus obtain as the average work per cycle
\beq
W_*\equiv \lim_{i\to\infty} \langle W_i^{\rm out}\rangle = \frac{2+t}{2(4+t)} \lim_{i\to\infty} \left( \yims-\yips \right) .
\label{eq:Ws}
\ee
The last limit is easily calculated by solving the dynamics (\ref{eq:y-dyn}) for the variance as
\beq
y^2(\tau)=1+e^{-2\tau}[y^2(0) -1]
\label{eq:y}
\ee
and setting $y^2(t) =(y_{i+1}^-)^2$ and $y^2(0)=\yips$. Using (\ref{eq:yips}),  and identifying $y_{i+1}^-$ with $y_i^-$ in
the limit $i\to\infty$, we get in the steady state for the variance before  a measurement the value
\begin{eqnarray}
({y^-_*})^2 \equiv \lim_{i\to\infty} \yims=\frac{1}{2} \bigg( &&1-y_m^2+e^{-2t}(y_m^2-1) \label{eq:messy_2}\\
&&+\sqrt{(y_m^2+1)^2-2e^{-2t}(y_m^4+1)+e^{-4t}(y_m^2-1)^2} \bigg)  \nonumber
\label{eq:yms}
\end{eqnarray}
with the limiting behavior
\begin{eqnarray}
(y_*^-)^2 \approx \left\{
\begin{array}{l l}
 y_m (2t)^{1/2} + (1-y_m^2)t  &{\rm ~~~for~~~} t\to 0  \\
 1 -\exp(-2t)/(1+y_m^2) &{\rm ~~~for~~~} t\to \infty .
\end{array}
\right.
\end{eqnarray}
Likewise, the variance after a measurement becomes
\beq
({y_*^+})^2 \equiv \lim_{i\to\infty} \yips=\frac{({y^-_*})^2 y_m^2}{({y^-_*})^2+y_m^2} 
\label{eq:yps}
\ee 
with the limiting behavior
\begin{eqnarray}
(y_*^+)^2 \approx \left\{
\begin{array}{l l}
 y_m(2t)^{1/2}-(1+y_m^2)t &  {\rm ~~~for~~~} t\to 0  \\
{\displaystyle \frac{y_m^2}{1+y_m^2} \left(1-\exp(-2t)\frac{y_m^2}{(1+y_m^2)^2}\right) } &{\rm ~~~for~~~} t\to \infty .
\end{array}
\right.
\end{eqnarray}
Finally, the average work per cycle
delivered by this  information machine
becomes
\beq
W_*=\frac{2+t}{2(4+t)} \left( (y^-_*)^2 -(y_*^+)^2 \right)
\ee
which is our first main result, shown in Fig. \ref{fig:w_k_fest}.
The power $P\equiv W_*/t$ becomes maximal if the cycle time becomes short with $P\approx 1/2-y_m(t/2)^{1/2}$ for
$t\to0$. In the long time limit,  $P\approx 1/[2t(1+y_m^2)]$ for $t\to\infty$.
 In the special case of an infinitely precise measurement, we get $P(t,0)=(1-e^{-2t})(2+t)/[2t(4+t)]$.

The efficiency of this machine follows from relating the power to the rate with which 
information is acquired through the 
measurements. The $i$-th measurement yields the information \cite{saga10}
\begin{eqnarray}
  \i_i =\int dx p(x|X_i)\ln [p(x|X_i)/p_i^-(x)],
\end{eqnarray}
which still depends on the result of all measurements $\{X_j\}_{1\leq j\leq i}$. By using (\ref{eq:pim}) and (\ref{eq:pxX}), subsequent averaging over the last measurement $X_i$ yields
\begin{eqnarray}
\I_{i}\equiv \int dX_i ~  p(X_i)\i_i= \ln(y_i^-/y_i^+).
\end{eqnarray}  
This simple result involves, a posteriori not surprisingly, just the variances before and after the measurement which are independent of the specific results $\{X_j\}_{1\leq j\leq i-1}$. $\I_i$ thus represents the information averaged over all measurement outcomes. In the stationary limit, one gets
\begin{eqnarray}
\I_*\equiv \ln(y^-_*/y^+_*)\approx \left\{
\begin{array}{l l}
  (t/2)^{1/2}/y_m -t/2 &{\rm ~~~for~~~} t\to 0  \\
{\displaystyle \frac{1}{2}\ln\left(1+\frac{1}{y_m^2}\right)-\frac{\exp(-2t)}{2(1+y_m^2)^2} } &{\rm ~~~for~~~} t\to \infty .
\end{array}
\right.
\end{eqnarray}
Consequently, the efficiency becomes
\beq
\eta\equiv W_*/\I_*=\frac{(2+t)\left( (y^-_*)^2 -(y_*^+)^2 \right)}{2(4+t)\ln(y_*^-/y_*^+)}
\ee
with the limiting behaviour
\begin{eqnarray}
\eta\approx \left\{
\begin{array}{l l}
 y_m(t/2)^{1/2} - y_m^2t/2 	& {\rm ~~~for~~~} t\to 0  \\
{\displaystyle \frac{1-2/t}{(1+y_m^2) \ln(1+1/y_m^2)} 	} &{\rm ~~~for~~~} t\to \infty .
\end{array}
\right.
\end{eqnarray}
As shown in Fig. \ref{fig:eta_k_fest}, the efficiency increases monotonically with the cycle time $t$. It becomes zero for $t\to 0$, which implies that this machine has vanishing efficiency at maximum power. The somewhat counterintuitive monotonic increase of $\eta$ with the measurement error $y_m$ arises from the fact that it is impossible to retrieve all information just by moving the center of the trap. Therefore, better measurements lead to a higher power but not to a higher efficiency. Indeed, while the work $W_*$ is bounded by $1/2$ \cite{abre11}, the information $\I_*$ diverges in the limit of infinitely precise measurements $y_m\to 0$, leading to a vanishing efficiency. In the limit $y_m\to \infty$, both $W_*$ and $\I_*$ tend to 0 and $\eta\to(2+t)/(4+t)$. For $t\to\infty$, this machine can reach the upper bound 1 imposed on $\eta$ by thermodynamics. However, this high efficiency is somewhat useless, since in this case the machine delivers vanishing power.

\begin{figure}[h]
 \centering
\includegraphics[width=0.45\textwidth]{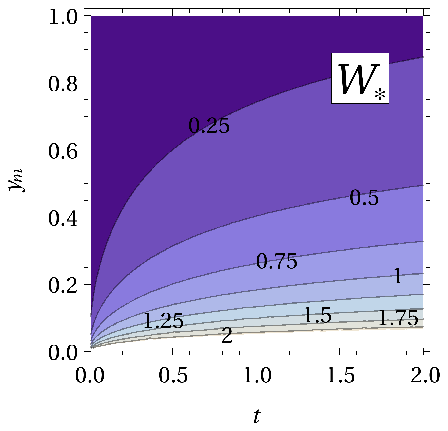}
\includegraphics[width=0.45\textwidth]{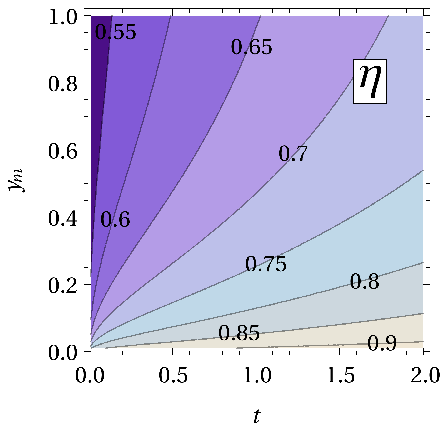}
\caption{Performance of the machine with optimally controlled stiffness $k(\tau)$ and center $\lambda(\tau)$.
Extracted work $W_*$ and efficiency $\eta$ both as function of cycle time $t$ and measurement error $y_m$.}
\label{fig:w_k_var}
\label{fig:eta_k_var}
\end{figure}

For a more powerful machine, we turn to a second variant where we allow additional control
over the stiffness of the trap $k(\tau)$. In this case, the contribution (\ref{eq:Wy}) no longer vanishes. It becomes maximal
for a  standard deviation $y(\tau)$ increasing linearly from $y(0)=y_i^+$ to
\beq
y(t)=y_{i+1}^-= \frac{y_i^+ + [(y_i^+)^2 + (2+t)t]^{1/2}}{2+t} .
\label{eq:yt}
\ee In the stationary limit, $i\to\infty$, using (\ref{eq:yt}) instead of (\ref{eq:y}) and the same reasoning to derive
the limiting behavior as above,
 we obtain for  the variance prior to a measurement in the steady state
$(y_*^-)^2$ the cubic equation
\beq
\left[(2+t)(y_*^-)^2-t\right]^2[(y_*^-)^2+ y_m^2]-4(y_*^-)^4 y_m^2= 0 
\label{eq:ys}
.\ee
The limiting behavior of its solution is
\begin{eqnarray}
(y_*^-)^2 \approx \left\{
\begin{array}{l l}
 y_m t^{1/2} + (3/4-y_m^2)t/2 & {\rm ~~~for~~~} t\to 0  \\
{\displaystyle 1 - \frac{2}{t} \left(1-\frac{y_m}{(1+y_m^2)^{1/2}}\right) } &{\rm ~~~for~~~} t\to \infty .
\end{array}
\right.
\end{eqnarray}
For (\ref{eq:yps}) one obtains $(y_*^+)^2$
with the short time and quasistatic behavior
\begin{eqnarray}
(y_*^+)^2 \approx \left\{
\begin{array}{l l}
 y_m t^{1/2} - (5/4+y_m^2)t/2 & {\rm for~} t\to 0  \\
{\displaystyle \frac{y_m^2}{1+y_m^2}\left(1-\frac{2 y_m^2 (1+y_m^2 -y_m (1+y_m^2)^{1/2})}{(1+y_m^2)^2 t} \right)}  &{\rm for~} t\to \infty .
\end{array}
\right.
\end{eqnarray}
For this second variant, we can still determine the contribution to the extracted work
from (\ref{eq:Wb}) as in the first case, provided we use the solution of (\ref{eq:ys}) in the expression
 (\ref{eq:Ws}) for the stationary limit.
Collecting everything, we obtain for the extracted work the expression
\beq
W_* =- \frac{ (y^-_*)^2 -(y_*^+)^2 }{4+t} - \frac{\left(y^-_*-y^+_* \right)^2}{t} +\ln(y^-_*/y^+_*)
\ee
shown in Fig. \ref{fig:w_k_var}.
In this case, the power diverges in the short limit as
\beq
P\equiv \frac{W_*}{t} \approx \frac{1}{4y_mt^{1/2}}
\ee
whereas in the long time limit one obtains
\beq
P \approx  \frac{1}{2t}\ln\left(1+\frac{1}{y_m^2}\right).\ee
The short time divergence of the power is compensated by a corresponding divergence of the
rate of information acquired through the measurements. Indeed, similarly as above, one gets the
information per measurement 
\begin{eqnarray}
\I_*\equiv \ln(y^-_*/y^+_*)\approx \left\{
\begin{array}{l l}
  t^{1/2}/(2y_m) -(4+1/y_m^2)t/16 & {\rm for~} t\to 0  \\
{\displaystyle  \frac{1}{2}\ln\left(1+\frac{1}{y_m^2}\right) - \frac{1-y_m/(1+y_m^2)^{\frac{1}{2}}}{t(1+y_m^2)} } & {\rm for~} t\to \infty .
\end{array}
\right.
\end{eqnarray}
The efficiency of this machine  becomes
\begin{eqnarray}
\eta &\equiv& W_*/\I_*=
1-\frac{ \left( (y^-_*)^2 -(y_*^+)^2 \right)/(4+t) +\left(y^-_*-y^+_* \right)^2/t}{\ln(y^-_*/y^+_*)}\\
&\approx& \left\{
\begin{array}{l l}
 1/2 + t^{1/2}/(8y_m) & {\rm ~~~for~~~} t\to 0  \\
1 - {\displaystyle \frac{4(1-y_m/(1+y_m^2)^{1/2})}{t\ln(1+1/y_m^2)} } & {\rm ~~~for~~~} t\to \infty ,
\end{array}
\right.
\end{eqnarray} shown in Fig. \ref{fig:eta_k_var}.

For this variant, the efficiency increases with the cycle time starting at $\eta=1/2$ for $t\to 0$
and saturating the upper bound $\eta =1$ for $t\to \infty$  and any $y_m$.
 In this quasistatic case, in contrast to the first variant, the two control parameters allow to extract the
full information. In another difference, the efficiency monotonically decreases with increasing $y_m$.
Here, more precise measurements  lead to a larger efficiency allowing even $\eta=1$ at finite $t$ for
infinite precision $y_m\to 0$. In the full $(t,y_m)$-plane, the efficiency
is bounded by 1/2 from below. The value 1/2 found here in the short time limit that corresponds to maximum power may hint to
a relation of our result with that for the efficiency of isothermal
machines at maximum power where the value 1/2 is universal in the linear response regime \cite{espo09}.
 While it is not obvious how to map repeated measurements for short cycle times
to a linear response formalism, finding the same value in both cases may be more than incidental.

In conclusion, we have studied the  efficiency
for a cyclically operating Brownian information machine
consisting of an overdamped particle in a time-dependent harmonic
trap. For two variants of such a machine, we have obtained analytically
how the  efficiency depends on both
the precision of a positional measurement and the cycle time.
Beyond these specific results 
our work raises a few questions concerning such machines
in general.
First, while
the quite natural definition of efficiency defined as mean extracted
work divided by the mean acquired information
shares features such as boundedness between  0 and 1 with the
more conventional thermodynamic definition of efficiency for
ordinary isothermal machines, finding $\eta = 1$  even for
finite cycle time in the limit of infinitely precise
measurements, as we do for the second variant, suggests that these 
information machines differ in essential aspects from 
thermodynamics ones. For reaching
$\eta=1$, the latter require a quasistatic operation, i.e.,
an infinite cycle time. Second, is it possible to formulate a 
linear response theory, i.e., to calculate Onsager coefficients
for such machines? Third, can we derive general
bounds on the efficiency at maximum power following 
reasoning for non-feedback driven machines?
Finally, an experimental test of such a machine would be interesting and 
should be possible with available technology.

\section*{References}
 \bibliographystyle{unsrt}
 \bibliography{/home/public/papers-softbio/bibtex/refs.bib}

\end{document}